\documentclass[twocolumn,10pts]{article}
\RequirePackage[left = 2cm, right = 2cm, top = 2.5cm, bottom = 3cm]{geometry}
\usepackage{amsmath} 
\usepackage{amsfonts}
\usepackage{amssymb}
\usepackage[parfill]{parskip}
\usepackage{booktabs}
\usepackage[utf8]{inputenc}
\usepackage{graphicx}
\usepackage{dcolumn}
\usepackage{bm}
\usepackage[usenames,dvipsnames,svgnames,table]{xcolor}
\usepackage{placeins}
\usepackage{caption} 
\usepackage{subcaption} 
\usepackage{cite}
\usepackage{authblk}
\usepackage{hyperref}
\usepackage{parskip}
\usepackage{titlesec}
\titleformat*{\section}{\bfseries}
\titleformat*{\subsection}{\itshape}

\newcommand{\acs}[1]{\langle\sigma_{#1}\, {\textrm v}\rangle}
\newcommand{\modu}[1]{\left\vert{#1}\right\vert}

\begin{document}

\title{\textcolor{RoyalPurple}{Gamma-ray line constraints on Coy Dark Matter}}

\author[1]{Andi Hektor}
 
\author[1,2]{Luca Marzola}

\author[1,2]{Taavi Tuvi}
 \affil[1]{Laboratory of High Energy and Computational Physics, National Institute of Chemical Physics and Biophysics, R\"avala pst. 10, 10143 Tallinn, Estonia. }
 \affil[2]{Laboratory of Theoretical Physics, Institute of Physics, University of Tartu; W. Ostwaldi tn 1, 50411 Tartu, Estonia.}
 
\twocolumn[
\begin{@twocolumnfalse}
\maketitle

\begin{abstract}
	\vskip1cm
	\noindent
Coy dark matter is an effective scheme in which a fermionic dark matter candidate interacts with the Standard Model fermions via a pseudoscalar mediator. This simple setup avoids the strong constraints posed by direct detection experiments in a natural way and explains, on top of the observed dark matter relic abundance, the spatially extended $\gamma$-ray excess recently detected at the Galactic Center. In this Letter we study the phenomenology of coy dark matter accounting for a novel signature of the model: the diphoton annihilation signal induced by the Standard Model fermions at the loop level. By challenging the model with the observations of spheroidal dwarf satellite galaxies and the results of $\gamma$-ray line searches obtained by the Fermi LAT experiment, we assess its compatibility with the measured dark matter relic abundance and the Galactic Center excesses. We show that despite the $\gamma$-ray line constraint rules out a significant fraction of the considered parameter space, the region connected to the observed Galactic Center excess remains currently viable. Nevertheless, we find that next-generation experiments such as DAMPE, HERD and GAMMA-400 have the potential to probe exhaustively this elusive scenario.
\vskip1cm
\end{abstract}
\end{@twocolumnfalse}]

\section{Introduction}

The matter content of our Universe is dominated by a component which, differently from ordinary matter, interacts at most very weakly with the photons of the Standard Model (SM) -- the dark matter (DM). It is usually assumed that DM consists of stable and weakly-interacting massive particles (WIMPs), which are thermal relics of dynamics once active in the hot early Universe. The reason behind the success of this picture is that particles with masses and annihilation cross sections set by the electroweak scale yield, in a natural way, DM relic densities of the order of the observed one. The basis of this remarkable coincidence lies in the freeze-out mechanism (for a review:~\cite{Jungman:1995df, Bertone:2004pz}), a natural consequence of the interplay between particle physics and an expanding Universe. For its simplicity and the appealing connection to frameworks like supersymmetry, the WIMP model became the paradigm of DM and shaped the dedicated long term experimental program. Direct detection experiments as XENON100~\cite{Aprile:2012nq}, PANDA~\cite{Tan:2016zwf} and LUX~\cite{Akerib:2016vxi} investigate the possible elastic scattering of DM on SM particles mediated by the postulated weak-scale interactions. Similarly, indirect detection investigations scrutinise potential traces of DM annihilations proceeding in the Galaxy or on cosmological scales via the same interactions. Finally, the direct production of DM particles in weak-scale phenomena is studied at collider experiments such as the LHC at CERN~\cite{Aad:2015zva,Khachatryan:2014rra}.

To date, in spite of the intense experimental effort, the nature and the properties of DM remain still a puzzle. Furthermore, the negative results of dedicated experiments accumulated so far have started to shake the belief of the community in the WIMP paradigm. In fact, the non-detection of supersymmetric partners of SM particles at collider experiments has impaired the attractiveness of supersymmetric theories, which traditionally provide a strong theoretical framework for WIMPs. Moreover, recent constraints from direct detection experiments started to challenged WIMP-nucleon scattering cross sections of the order of the typical weak-scale value, thereby excluding substantial parts of the parameter spaces of traditional WIMP models. On the other hand, during the last decade, indirect detection experiments have reported anomalies that could be a first manifestation of DM. In 2008 the PAMELA satellite measured an excess of cosmic positrons above the energy of 20~GeV that could be a byproduct of DM annihilation/decay in our Galaxy~(e.g.~\cite{Cirelli:2008pk, ArkaniHamed:2008qn}). The signal was later confirmed by subsequent analyses from the Fermi LAT and AMS-02 collaborations~\cite{Adriani:2008zr, Abdo:2009zk, Aguilar:2013qda}. Similarly, in 2009 the Fermi LAT data~\cite{Atwood:2009ez} revealed a spatially extended $\gamma$-ray excess at the Galactic Centre (GC) in the energy windows of 1-5~GeV~\cite{Goodenough:2009gk, Hooper:2010mq, Abazajian:2010zy, Boyarsky:2010dr, Hooper:2011ti, Abazajian:2012pn, Gordon:2013vta, Macias:2013vya, Abazajian:2014fta, Daylan:2014rsa, Lacroix:2014eea}, whereas in 2012 hints of a $\gamma$-ray line at 130~GeV were found~\cite{Bringmann:2012vr, Weniger:2012tx, Tempel:2012ey, Su:2012ft} in the updated data. Intriguingly, these features all point to DM annihilation cross sections of the order of the freeze-out one in WIMP models, $\acs{ann} \simeq 3\times 10^{-26}$~cm$^3$s$^{-1}$, supporting the fascinating idea that DM interactions could originate the observed signals. However, we remind that the environment where such phenomenon are observed,  mainly the GC, is an extremely complex region filled with stars, stellar relics, dust, gas, and subject to intense cosmic rays. As active astrophysical processes like millisecond pulsar populations~\cite{Yuan:2014rca} or ultra-energetic events from the past~\cite{Petrovic:2014uda} also can explain the mentioned observations, it is currently not possible to discern their origin.

In order to investigate whether DM dynamics is behind the mentioned signals, we explore here complementary phenomenological implications within the framework of ``coy dark matter'' (CoyDM)~\cite{Boehm:2014hva}: a model that draws from the WIMP paradigm but eludes the stringent direct detection bounds in a natural way. In this effective scheme, DM is a new Dirac fermion that interacts with the SM content by the exchange of a pseudoscalar mediator. DM can then annihilate into SM fermions to give rise to the observed DM relic abundance, via the freeze-out mechanism, and to secondary photon signals in DM dense regions such as the GC or dwarf satellite galaxies. Interestingly, owing to the pseudoscalar nature of the mediator, the scattering of CoyDM on nucleons is here a spin-dependent process. The corresponding direct detection constraints are consequently orders of magnitudes weaker than the spin-independent bounds that apply to traditional WIMP scenarios, and allow the CoyDM model to match the observed signals without fine tunings in its parameters~\cite{Boehm:2014hva}. Given the flexibility of the resulting framework, CoyDM was even proposed as a solutions to other open problems in particle physics, for instance the annual modulation observed in the DAMA/LIBRA experiment~\cite{Arina:2014yna} or the measured anomalous magnetic moment of muon~\cite{Hektor:2014kga}.

In the attempt to bound the properties of this elusive framework, in this Letter we detail a yet unexplored feature of coy dark matter; the diphoton signal originated at the loop level by its interactions with the SM fermions.  DM annihilations resulting into a diphoton final state induce line features in the cosmic $\gamma$-ray spectrum, which are easily distinguished from the power-law background due to astrophysical processes. Motivated by this observation, we explore the parameter space of CoyDM by challenging the model with: (i) the observed DM relic abundance, (ii) the constraints imposed by observations of dwarf spheroidal satellites ~\cite{Ackermann:2015zua}, (iii) the bounds resulting from $\gamma$-ray line searches in the GC region~\cite{Ackermann:2015lka} and (iv) the broad photon excess detected at the GC in an energy window of $1\div5$ GeV~\cite{Goodenough:2009gk, Hooper:2010mq, Abazajian:2010zy, Boyarsky:2010dr, Hooper:2011ti, Abazajian:2012pn, Gordon:2013vta, Macias:2013vya, Abazajian:2014fta, Daylan:2014rsa, Lacroix:2014eea}\footnote{ Although we focus here on the constraints posed by the measurements of the cosmic photon spectrum, we remark that CoyDM has the potential to explain $\gamma$-ray lines possibly detected in the cosmic spectrum, for instance the $\sim$40~GeV excess reported in Ref.~\cite{Liang:2016pvm}. The same excess is also shown at the GC  and galaxy clusters by the other studies, albeit at a lower significance~\cite{Ackermann:2015lka, Anderson:2015dpc}. As we will demonstrate, the CoyDM is able to reproduce such signals owing to annihilations via the $\gamma\gamma$ final state, induced at the loop level by the charged fermions of the SM (and beyond SM).}. In the context of $\gamma$-ray line searches, we also evaluate the reach of forthcoming experiments such as DAMPE~\cite{Li:2012qg, Lu:2013kda, ChangJin:550, Gargano:2017avj}, HERD~\cite{Zhang:2014qga} and GAMMA-400~\cite{2013AdSpR..51..297G}, showing the region of parameter space that these could probe. With our analysis we show that the $\gamma$-ray line searches based on the Fermi LAT data significantly bound the properties of CoyDM, although the current constraints are not able to probe the region of the	 parameter space associated to the detected GC excess. Remarkably, we find instead that the next-generation experiments have the capability to exhaustively explore the considered parameter region. In fact, these experiments can potentially preclude CoyDM from explaining the observed GC excess, and even relegate the model to narrow corners of its parameter space where the DM relic abundance bound is matched only owing to resonance effects.

The paper is organised as follows. In Sec.~\ref{sec:Coy Dark Matter} we briefly review the model and introduce the cross sections relevant for our analysis, highlighting the diphoton one. The details of our investigation and the considered experimental constraints are specified in Sec.~\ref{sec:bounds}, where we show the impact of the mentioned bounds on the parameter space of CoyDM. Finally, in Sec.~\ref{sec:Conclusions} we draw our conclusions.

\section{Coy Dark Matter} 
\label{sec:Coy Dark Matter}

Following the proposal of Boehm~et~al~\cite{Boehm:2014hva}, our DM candidate is a Dirac fermion $\chi$ with mass $m_{\chi}$. The interactions of $\chi$ with the SM content are mediated by a new pseudoscalar field $a$, of mass $m_a$, according to the effective Lagrangian
\begin{equation}\label{eq:lag}
{\cal L} = {\cal L}_{SM} - i\frac{g_{\chi}}{\sqrt 2} \, a \bar\chi \gamma^5 \chi - i\sum_{f} \frac{g_f}{\sqrt 2} \, a \bar f \gamma^5 f + \text{ H.c.}
\end{equation} 
where $f$ runs on the SM fermions and
\begin{equation}\label{eq:defA}
g_f := A_f \, y_f = A_f\,\frac{\sqrt 2 \, m_f}{v} 
\end{equation}
being $v = 246$~GeV the Higgs boson vacuum expectation value. 
The couplings of the pseudoscalar mediator to the SM fermions are assumed  proportional to the corresponding Higgs Yukawa couplings, in agreement with the minimal flavour violation ansatz~\cite{DAmbrosio:2002ex}. In the following analysis we will set the proportionality factor $A_f=1$, commenting however on the impact of a different choice on our results.

In this setup, the DM annihilation rate into SM fermions at the present era amounts to 
\begin{equation}\label{eq:bb}
	\acs{\chi\chi \to f\bar f}_0 
	\simeq 
	\frac{C_f (g_\chi g_f)^2}{8\pi} 
	\frac{m_\chi^2 \sqrt{1-m_f^2/{m_\chi^2}}}
	{(m_a^2-4m_\chi^2)^2+m_a^2 \Gamma_a^2}
\end{equation}
where $C_f$ is the color multiplicity of a SM fermion $f$ and, as customary, we approximated the thermal average by retaining the $s$-wave contribution only. In the above formula we indicated with $\Gamma_a$ the total decay width of $a$, given by
\begin{equation}
	\Gamma_a = \sum_{\bar f}\frac{C_f (g_\chi y_{\bar f})^2 \, m_a}{16\pi} \sqrt{1-\frac{4 m_{\bar f}^2}{m_a^2}},
\end{equation}
where $\bar f$ runs on fermions, DM included, with masses $m_{\bar f}<m_a/2$ and we neglected higher order contributions which allow the pseudoscalar mediator to decay into the SM gauge bosons. 

The above cross section regulates the intensity of the potential signal from DM annihilations emitted at the GC and dwarf spheroidal satellites of the Milky Way. In the following we will also consider the results of $\gamma$-ray line searches obtained by the Fermi LAT collaboration, which bound the loop level $\gamma\gamma$-channel annihilation rate,
\begin{align}\label{eq:gamma-line}
	\acs{\chi\chi \to \gamma\gamma} &= \frac{g_\chi^2\alpha^2}{16\pi^3}
\frac{\modu{\sum_f C_f Q^2_f g_f \, m_f {\cal F}\left(\frac{m_\chi^2}{ m^2_f}\right)}^2}{(4m_\chi^2-m^2_a)^2+m^2_a\Gamma_a^2}.
\end{align}
Here $f$ is running on the fermions in the loop that connects the pseudoscalar mediator to the SM photons, while $Q_f$, $m_f$ and $C_f$ are respectively the electric charge, mass and colour multiplicity of these particles. The loop function that enters the above expression is given by~\cite{Djouadi:2005gi,Djouadi:2005gj}
\begin{equation}
{\cal F} = \begin{cases}
\arcsin^2(\sqrt x) &\text{if }x\leq1\\
-\dfrac{1}{4}\left[\log\left(\frac{1+\sqrt{1-x^{-1}}}{1-\sqrt{1-x^{-1}}}\right)-i\pi\right]^2 &\text{if }x>1
\end{cases}.
\end{equation} 
We find that the large contributions from the light quarks are compensated by the smallness of the associated couplings $g_f \propto m_f / v$. Consequently the annihilation rate of DM into photons is dominated by the contributions of heavy quarks running in the loop. 

For the chosen set of pseudoscalar-SM couplings, the $\gamma$-ray line cross-section is typically suppressed by a few orders of magnitude with respect to the $f \bar f$ one. However, we remark that line signals in the galactic photon spectrum can be distinguished from the power-law astrophysical background more easily than distributed excesses due to primary fermionic annihilation channels. 

\section{Indirect Detection and relic abundance bounds} 
\label{sec:bounds}
The model, as delineated in the previous section, is completely specified by three parameters: $m_a$, $m_\chi$ and $g_\chi$. Our exploration of the corresponding parameter space focuses on the ranges reported in table~\ref{tab1}. We explore a range of mediator masses complementary to that probed by flavour physics constraints~\cite{Dolan:2014ska}, whereas the values of DM mass that we consider are motivated by the phenomenology of the GC excess.
 
\setlength{\tabcolsep}{2pt}
\begin{table}[htb!]
	\centering
	\begin{tabular}{c|c|c|c}
	\toprule
	\textbf{Parameter} & \textbf{Lower bound} & \textbf{Upper bound}& \textbf{Step}\tabularnewline
	\midrule
	$m_a$ (GeV) & 5 & 300 & 5\tabularnewline
	$m_\chi$ (GeV) & 20 & 120 & 1\tabularnewline
	$g_\chi$ & $10^{-2}$ & 10 & $10^{-3}$\tabularnewline
	\bottomrule
	\end{tabular}
	\caption{The ranges considered for the CoyDM parameters in the performed scan. We set $A_f=1$, cf. eq.~\eqref{eq:defA}.}
	\label{tab1}
\end{table}

\begin{figure*}[h]
	\includegraphics[width=0.32\textwidth]{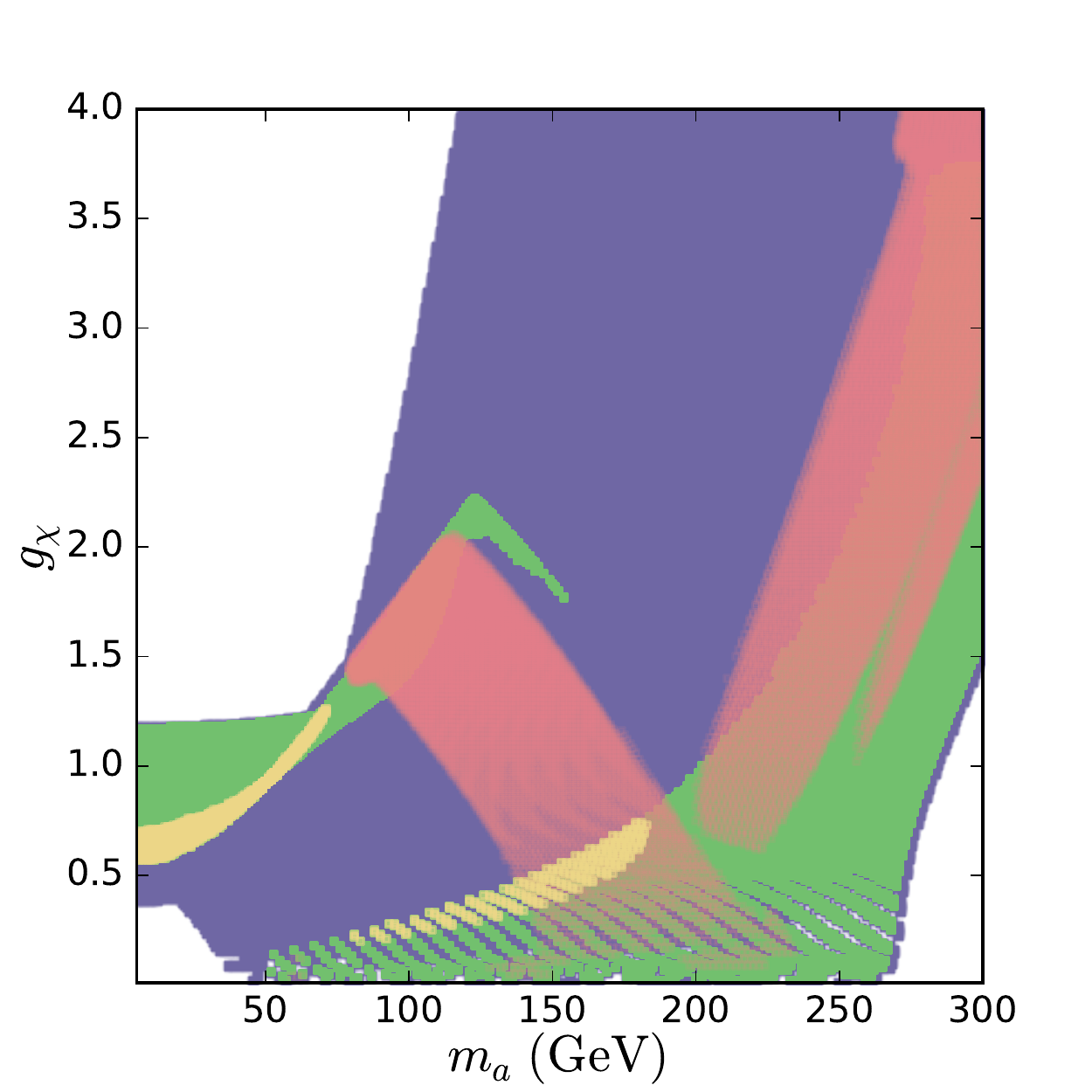}
	\includegraphics[width=0.32\textwidth]{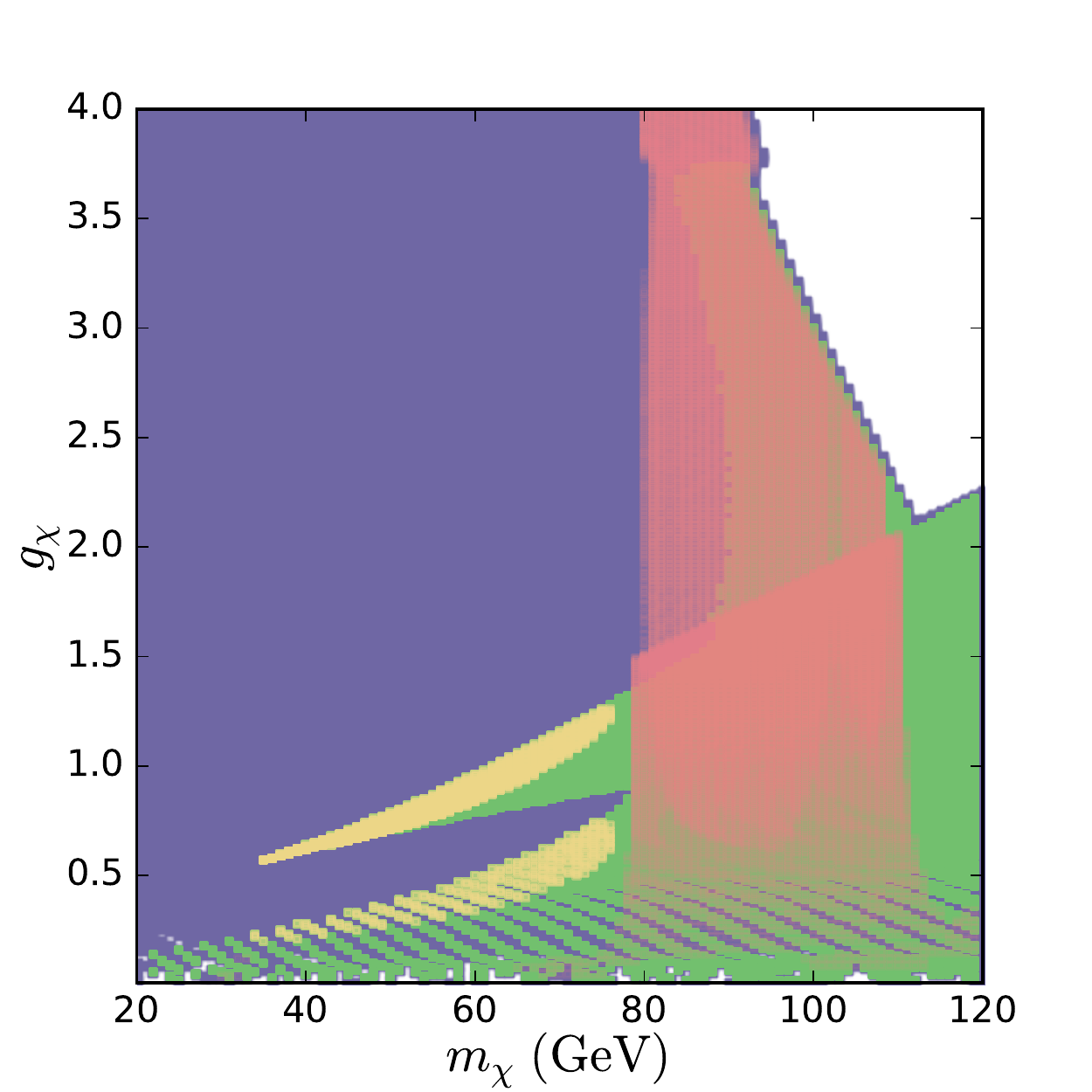}
	\includegraphics[width=0.32\textwidth]{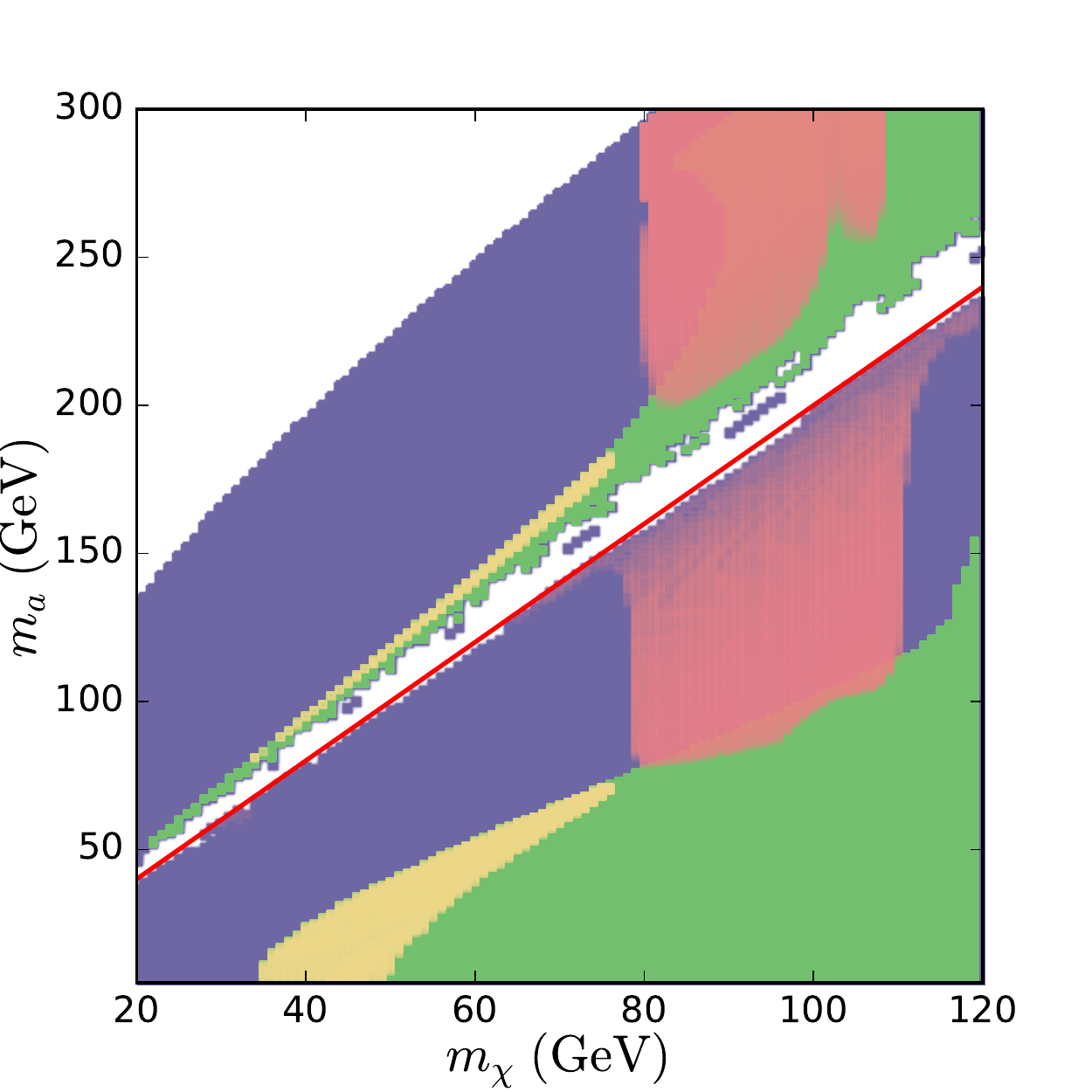}
	\caption{The explored parameter space of the CoyDM model. The blue region indicates the area where the observed DM relic abundance is matched. The green areas, instead, single out the parameters of the model which satisfy the bound from dwarf satellite galaxies on top of the DM relic abundance. In the yellow regions, as well as obeying the previous constraint, the model is able to reproduce the broad photon excess detected at the Galactic Center. The red areas, instead, are excluded by the non-observation of $\gamma$-ray lines in the photon spectrum detected by the Fermi LAT satellite (observed limit~\cite{Ackermann:2015lka}).}
	\label{fig1}
\end{figure*}

In this regard, previous analysis of the GC excess~\cite{Gordon:2013vta, Macias:2013vya, Abazajian:2014fta, Daylan:2014rsa, Calore:2014nla} found that the detected photon signal is best reproduced by DM annihilation proceeding via the bottom-quark channel. In the considered CoyDM model, if the mass of the DM particle $\chi$ is below the mass of the top quark, the dominant annihilation channel is precisely $\chi\chi \to b\bar{b}$ because of the hierarchy in the pseudoscalar couplings to the SM fermions. In particular, the choice $m_\chi \simeq 30$~GeV allows to fit the $\gamma$-ray excess for a natural value of the DM annihilation cross section $\sim\acs{ann}\simeq 3\times 10^{-26}$ cm$^3$s$^{-1}$. 

We remark that the excess can also be reproduced via the $\chi\chi \to \tau\bar{\tau}$ channel, see for instance ~\cite{Hektor:2014kga, Hektor:2015zba}, although in this case a lower DM mass $m_\chi \simeq 10$~GeV is required and the quality of the fit worsens with respect to the $b\bar{b}$-channel result. 

Having delineated the region of the parameter space that our analysis explores, we detail below the constraints that we consider. 

\begin{itemize}
	\vskip\baselineskip
\item \underline{DM relic abundance}\\
For every point in the parameter space we check whether the current constraint on the DM relic abundance is satisfied. This quantity is computed through the dedicated function provided by the \texttt{micrOMEGAs v.4.3.2} library~\cite{Belanger:2010pz,Belanger:2013oya}. We consider a point successful if it yields a relic abundance $\Omega_{DM}\,h^2\in[0.1118, 0.1199]$, corresponding to the $3\sigma$ bound from the Planck experiment~\cite{Ade:2015xua}. 
\vskip\baselineskip
\item \underline{GC excess}\\
In this study we adopt the best-fit values obtained in ref.~\cite{Calore:2014nla} for the cross-section $\acs{\chi\chi \to b\bar b}$; a point successfully reproduces the GC excess if according to eq.~\eqref{eq:bb} it leads to a $b \bar b$-channel annihilation cross section within $3\sigma$ from the best fit value of ref.~\cite{Calore:2014nla}.
\vskip\baselineskip
\item \underline{Bound from observations of dwarf galaxies}\\
Observations of the dwarf spheroidal satellites of the Milky Way by the Fermi LAT pose a stringent bound on the DM annihilation rate for the quark and tauonic channels. In our analysis we require that $\acs{\chi\chi \to b\bar b}$ fall within the $95\%$ confidence interval from ref.~\cite{Ackermann:2015zua}.
\vskip\baselineskip
\item \underline{The $\boldsymbol{\gamma}$-ray line constraint }\\
As mentioned before, despite the reduced cross section, $\gamma$-ray lines are easily recognisable over the power-law spectrum of astrophysical background. For this reason we check that the values obtained for the diphoton cross section via eq.~\eqref{eq:gamma-line} comply with the observed bound from ref.~\cite{Ackermann:2015lka}. We also consider the impact of the DAMPE~\cite{Gargano:2017avj,Lu:2013kda,Li:2012qg}, HERD~\cite{Zhang:2014qga} and GAMMA-400~\cite{2013AdSpR..51..297G} next generation experiments by estimating their reach from refs.~\cite{Li:2012qg,Bergstrom:2012vd,2013AIPC.1516..288G,Lu:2013kda,Charles:2016pgz,Gargano:2017avj}. 
\end{itemize}

The scan plots in figure~\ref{fig1} show projections of the parameter space of the model in the considered ranges and the effect of the mentioned constraints. We highlight in blue the regions of the parameter space which result in the correct DM relic abundance. Although the required values are achieved in most of the considered configurations, the strict bound posed by observations of dwarf spheroidal satellite galaxies restrict considerably the viable parameter space~\cite{Ackermann:2015zua}. We highlight in green the regions where both the bound on the annihilation cross section from dwarf spheroidal satellites and that on the DM relic abundance are satisfied. The red regions, instead, delineate the areas that are excluded by the statistical constraint from the non-observation of $\gamma$-ray lines in the cosmic photon spectrum~\cite{Ackermann:2015lka} by the Fermi LAT experiment. Finally, in the yellow regions, the model satisfies the DM abundance constraint, the constraint on $\acs{\chi\chi \to b\bar b}$ due to observations of the dwarf satellite galaxies and reproduces, as well, the broad photon excess observed at the GC~\cite{Gordon:2013vta, Macias:2013vya, Abazajian:2014fta, Daylan:2014rsa, Calore:2014nla}. We signaled with a red line in the rightmost panel the points that satisfy the resonance condition $m_a=2m_\chi$. Neighbouring configurations are excluded because of the large resonant enhancement to the total annihilation cross section. 

\begin{figure}[h]
	\centering
	\includegraphics[width=0.49\textwidth]{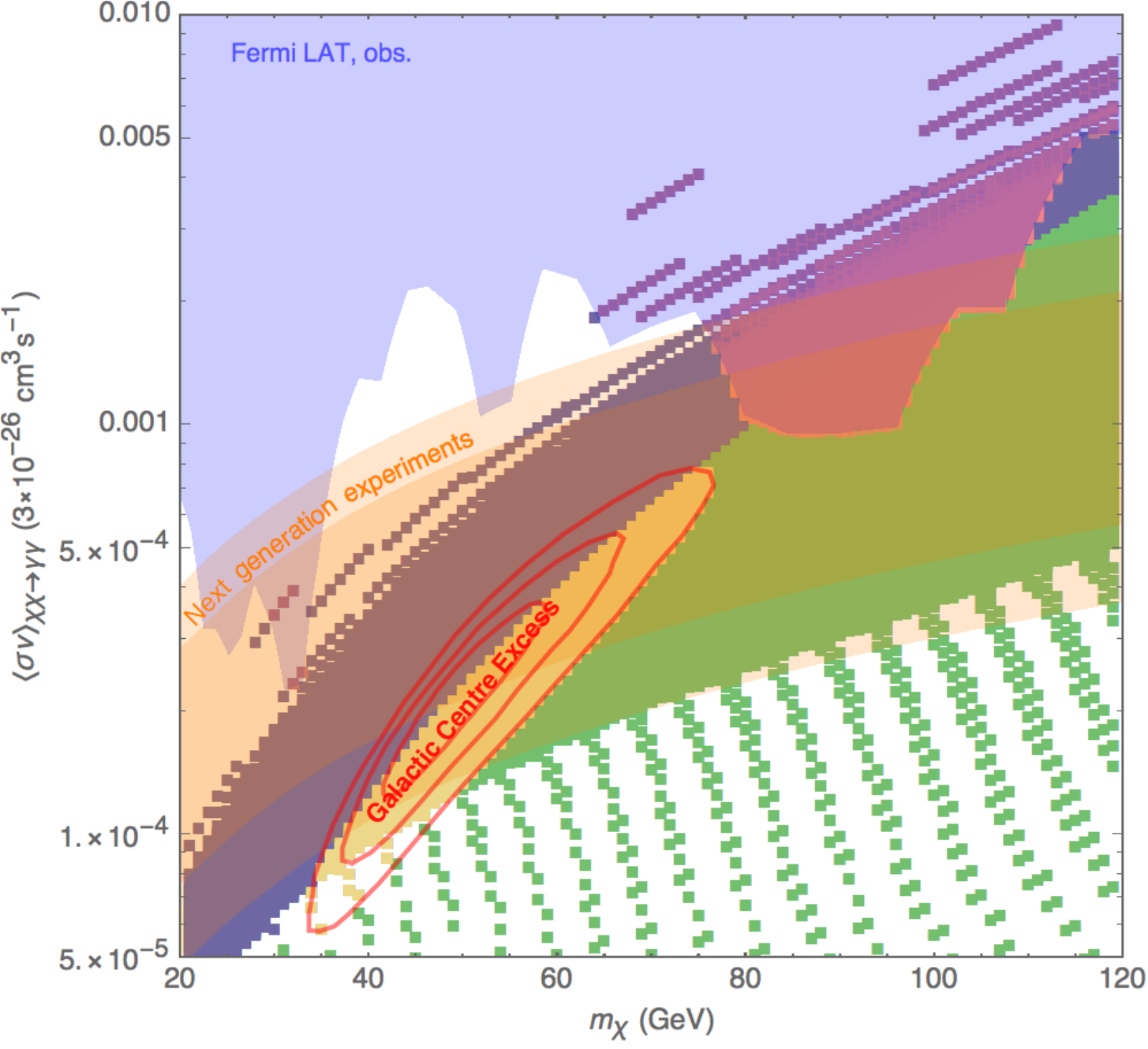}
	\caption{Constraints posed by the $\gamma$-ray line searches on the parameter space of the coy dark matter model. The adopted color code is the same as in figure~\ref{fig1}. The red ovals highlight the regions where the model fits the broad photon excess at the Galactic Center with a confidence level of 1, 2 and 3$\sigma$. The light blue region shows instead the current $95\%$ confidence level bound from the Fermi LAT $\gamma$-ray line searches, which excludes the region of the parameter space shaded in semi-transparent red. The orange bands show our estimates for the $68\%$ (darker) and $95\%$ (lighter) containment limits of the exclusion bound that next-generation experiments will cast.}
	\label{fig2}
\end{figure}

The constraints posed by the $\gamma$-ray line searches are exposed in greater detail in figure~\ref{fig2}. Here we plot the relevant annihilation rate $\acs{\chi\chi\to \gamma\gamma}$ as a function of the DM mass $m_\chi$, adopting the same color code as in figure~\ref{fig1}. The red ovals highlight the 1, 2 and 3$\sigma$ confidence intervals for the GC excess fit, whereas the light blue regions represents the observed $95\%$ confidence level limit from the Fermi LAT $\gamma$-ray line searches. As mentioned before, the  areas in semi-transparent red are excluded by this constraint at a corresponding confidence level.

The reach of next generation experiments ~\cite{Gargano:2017avj,Lu:2013kda,Li:2012qg,Zhang:2014qga,2013AdSpR..51..297G} is indicated by the orange bands, which represent our estimate based on the specifications in refs.~\cite{Li:2012qg,Bergstrom:2012vd,2013AIPC.1516..288G,Lu:2013kda,Charles:2016pgz,Gargano:2017avj} for the $95\%$ (lighter) and $68\%$ (darker) containment limits of the expected exclusion bound. As we can see, for their improved energy resolution, these experiments can probe a vast majority of the considered parameter space, covering in particular almost the totality of the region where the CoyDM model reproduces the GC photon excess. The possibility that the mentioned future observations could provide such an exhaustive test of this elusive model is indeed intriguing. Notice also that the next generation experiments could reach diphoton cross sections below about $3\times 10^{-30}$ cm$^3$ s$^{-1}$, forcing the CoyDM model to satisfy the DM relic abundance constraint only through resonance effects. The latter result in the periodic pattern shown in the bottom part of the figure, the spacing of which reflects the values of the adopted increment steps, see table~\ref{tab1}. Finally, we remark that the bound cast by $\gamma$-ray line searches is also strengthened in extension of this framework that contain new charged heavy states coupled to the pseudoscalar mediator.
 
\begin{figure}[h]
	\centering
	\includegraphics[width=0.45\textwidth]{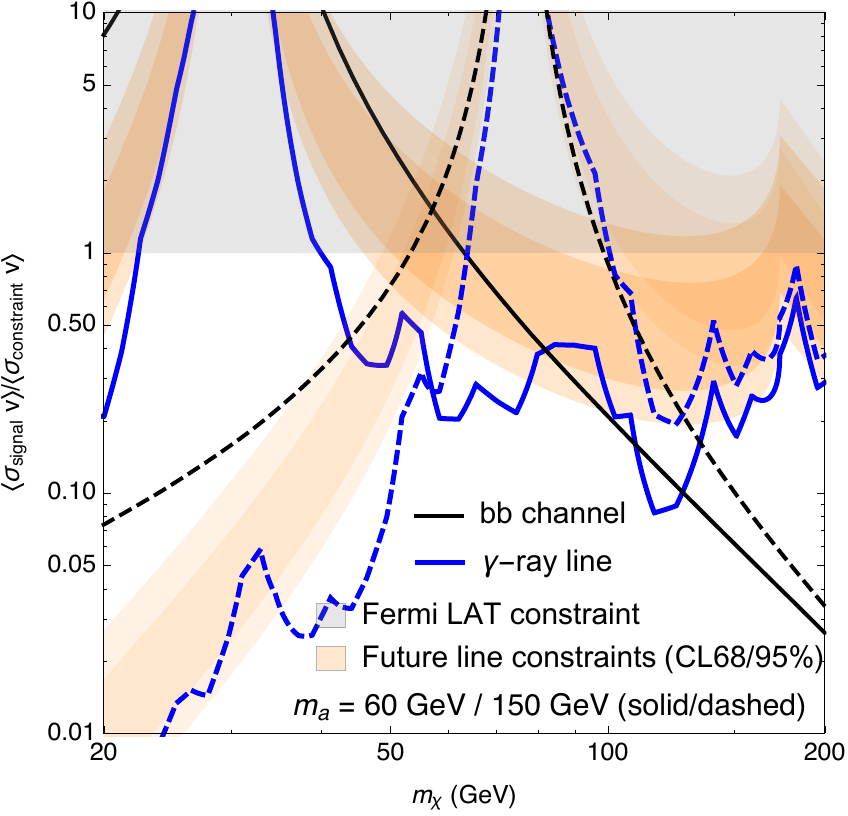}
	\caption{Sensitivity of the constraints from $\gamma$-ray line searches and observations of dwarf spheroidal satellites to the enhancement factor $A_f$ in eq.~\eqref{eq:defA}. The blue and black lines show the ratios of potential signal annihilation rates with respect to the corresponding bounds, respectively for the $\gamma\gamma$ and $b\bar{b}$ channel. The solid and dashed lines are for two different mediator mass, respectively $m_a = 60$ GeV and $m_a = 150$ GeV. The light gray shading denotes the experimental exclusion area, where the signal annihilation rates overshoot their bounds. The orange bands denote instead our estimate for the $68\%$ and $95\%$ containment regions of next-generation experiments. }
	\label{fig3}
\end{figure}

To conclude our analysis, we investigate the impact that larger values of the enhancement factor $A_f$ in eq.~\eqref{eq:defA} have on the analysed bounds. Given that both the annihilation rates in eq.~\eqref{eq:bb} and~\eqref{eq:gamma-line} scale as $\acs{f\bar{f}\to b\bar{b}/\gamma\gamma}\propto A_f^2$, we compute the ratios between these quantities and the corresponding experimental constraints, cast respectively by observation of dwarf spheroid satellites~\cite{Ackermann:2015zua} and $\gamma$-ray line searches~\cite{Ackermann:2015lka}. The ratios are illustrated in figure~\ref{fig3}, where the case of $b \bar{b}$-channel is plotted in black whereas the diphoton one is shown in blue. The solid and dashed lines represent different choices of the mediator mass: $m_a = 60$ GeV and $m_a = 150$ GeV respectively. We shaded in gray the region of the plot where the considered annihilation rates overshoot the corresponding constraints, being thereby excluded at progressive significances by such observations. The constraint on the $b\bar{b}$ annihilation channel is tighter than the $\gamma$-ray line one on most of the considered DM mass range, independently of the chosen mediator mass $m_a$. For larger values of the enhancement factor $A_f$ and a fixed mediator mass, the constraint of dwarf spheroidal satellites has the potential to rule out areas in the parameter space characterised by $m_\chi\lesssim 100$ GeV. For higher DM masses, the $\gamma$-ray line searches are generally expected to be more sensitive to enhancements in the the pseudoscalar mediator couplings to the SM fermions. In figure~\ref{fig3} we also show the estimated $68\%$ and $95\%$ containment bands for the considered next-generation experiments, respectively shaded in dark and light orange. The experiments will probe approximately the same DM mass range as the current searches, although at a much higher sensitivity.

\section{Results} 
\label{sec:Conclusions}

The coy dark matter model reconciles dynamics typical of the WIMP paradigm with the latest results of direct detection experiments. Owing to the pseudoscalar nature of the mediator, that bridges here the dark and visible sectors, the coy dark matter model naturally evades the stringent bounds on the spin-independent cross section associated to elastic dark matter-nucleon scatterings. Current measurements of the corresponding spin-dependent cross section, which in principle could constrain the model, are not yet sensitive enough to cast effective bounds. 

In order to explore the coy dark matter scenario, in this Letter we have investigated the related phenomenology in the context of indirect detection experiments, with particular attention to possible $\gamma$-ray lines signatures. More in detail, we scanned a sensible region of the associated parameter space assessing the impact of the considered bounds: the observed dark matter abundance, the broad $\gamma$-ray excess detected at the Galactic Center, the $\gamma$-ray observations of the dwarf spheroidal satellite galaxies of the Milky Way and the searches for $\gamma$-ray lines in the Galactic photon spectrum by the Fermi LAT. The last observable has been discussed also in relation to the upcoming DAMPE, HERD and GAMMA-400 experiments.

Our results can be summarised as follows:

\begin{itemize}
 
 \item The coy dark matter model is able to give rise to the observed dark matter relic abundance in a vast part of the considered parameter space via the freeze-out mechanism, implemented here predominantly by dark matter annihilation to the $b \bar b$ final state. However, the observational $\gamma$-ray bounds from the dwarf spheroidal satellites limit noticeably the number of successful solutions.
 
 \item The mentioned bounds do not preclude the model from fitting the $1\div5$~GeV $\gamma$-ray excess detected at the Galactic Center via the same $b \bar b$ final state. 
 
 \item The constraints cast by $\gamma$-ray line searches started to probe the scenario, excluding a sizeable area of the parameter space for dark matter masses larger than about 80 GeV. Unfortunately, the current data is not able to test the regions where the coy dark matter model can explain the Galactic Center excess.
 
 \item The above-listed next-generation $\gamma$-ray experiments have the potential to perform an exhaustive test of the scenario. Our estimate of the $68\%$ and $95\%$ containment bands for the expected exclusion limit cover almost the totality of the considered parameter space, including the region associated to the $\gamma$-signal detected at the Galactic Center. It is possible that next-generation experiments will corner the model to narrow regions of its parameter space where the DM relic abundance constraint is satisfied only via resonance effects in the dark matter annihilation process.
\end{itemize}

\section*{Acknowledgements}

The authors acknowledge the Estonian Research Council for supporting their work with the grants PUT808 and PUTJD110. AH thanks the Horizon 2020 programme as this project has received funding from the European Union’s Horizon 2020 program under the Marie Sklodowska-Curie grant agreement No 661103. This work was also supported by the European Social Fund grants IUT23-6, CERN+ and through the ERDF CoE grant TK133.

\bibliographystyle{JHEP}
\bibliography{./coy.bib}

\providecommand{\href}[2]{#2}\begingroup\raggedright\begin{thebibliography}{10}

\bibitem{Jungman:1995df}
G.~Jungman, M.~Kamionkowski, and K.~Griest, {\it {Supersymmetric dark matter}},
   {\em Phys.Rept.} {\bf 267} (1996) 195--373,
  [\href{http://arxiv.org/abs/hep-ph/9506380}{{\tt hep-ph/9506380}}].

\bibitem{Bertone:2004pz}
G.~Bertone, D.~Hooper, and J.~Silk, {\it {Particle dark matter: Evidence,
  candidates and constraints}},  {\em Phys.Rept.} {\bf 405} (2005) 279--390,
  [\href{http://arxiv.org/abs/hep-ph/0404175}{{\tt hep-ph/0404175}}].

\bibitem{Aprile:2012nq}
{\bf XENON100} Collaboration, E.~Aprile et~al., {\it {Dark Matter Results from
  225 Live Days of XENON100 Data}},  {\em Phys. Rev. Lett.} {\bf 109} (2012)
  181301, [\href{http://arxiv.org/abs/1207.5988}{{\tt arXiv:1207.5988}}].

\bibitem{Tan:2016zwf}
{\bf PandaX-II} Collaboration, A.~Tan et~al., {\it {Dark Matter Results from
  First 98.7-day Data of PandaX-II Experiment}},
  \href{http://arxiv.org/abs/1607.07400}{{\tt arXiv:1607.07400}}.

\bibitem{Akerib:2016vxi}
D.~S. Akerib et~al., {\it {Results from a search for dark matter in LUX with
  332 live days of exposure}},  \href{http://arxiv.org/abs/1608.07648}{{\tt
  arXiv:1608.07648}}.

\bibitem{Aad:2015zva}
{\bf ATLAS} Collaboration, G.~Aad et~al., {\it {Search for new phenomena in
  final states with an energetic jet and large missing transverse momentum in
  pp collisions at $\sqrt{s}=$8 TeV with the ATLAS detector}},  {\em Eur. Phys.
  J.} {\bf C75} (2015), no.~7 299, [\href{http://arxiv.org/abs/1502.01518}{{\tt
  arXiv:1502.01518}}]. [Erratum: Eur. Phys. J.C75,no.9,408(2015)].

\bibitem{Khachatryan:2014rra}
{\bf CMS} Collaboration, V.~Khachatryan et~al., {\it {Search for dark matter,
  extra dimensions, and unparticles in monojet events in proton--proton
  collisions at $\sqrt{s} = 8$ TeV}},  {\em Eur. Phys. J.} {\bf C75} (2015),
  no.~5 235, [\href{http://arxiv.org/abs/1408.3583}{{\tt arXiv:1408.3583}}].

\bibitem{Cirelli:2008pk}
M.~Cirelli, M.~Kadastik, M.~Raidal, and A.~Strumia, {\it {Model-independent
  implications of the e+-, anti-proton cosmic ray spectra on properties of Dark
  Matter}},  {\em Nucl. Phys.} {\bf B813} (2009) 1--21,
  [\href{http://arxiv.org/abs/0809.2409}{{\tt arXiv:0809.2409}}]. [Addendum:
  Nucl. Phys.B873,530(2013)].

\bibitem{ArkaniHamed:2008qn}
N.~Arkani-Hamed, D.~P. Finkbeiner, T.~R. Slatyer, and N.~Weiner, {\it {A Theory
  of Dark Matter}},  {\em Phys. Rev.} {\bf D79} (2009) 015014,
  [\href{http://arxiv.org/abs/0810.0713}{{\tt arXiv:0810.0713}}].

\bibitem{Adriani:2008zr}
{\bf PAMELA} Collaboration, O.~Adriani et~al., {\it {An anomalous positron
  abundance in cosmic rays with energies 1.5-100 GeV}},  {\em Nature} {\bf 458}
  (2009) 607--609, [\href{http://arxiv.org/abs/0810.4995}{{\tt
  arXiv:0810.4995}}].

\bibitem{Abdo:2009zk}
{\bf Fermi-LAT} Collaboration, A.~A. Abdo et~al., {\it {Measurement of the
  Cosmic Ray e+ plus e- spectrum from 20 GeV to 1 TeV with the Fermi Large Area
  Telescope}},  {\em Phys. Rev. Lett.} {\bf 102} (2009) 181101,
  [\href{http://arxiv.org/abs/0905.0025}{{\tt arXiv:0905.0025}}].

\bibitem{Aguilar:2013qda}
{\bf AMS} Collaboration, M.~Aguilar et~al., {\it {First Result from the Alpha
  Magnetic Spectrometer on the International Space Station: Precision
  Measurement of the Positron Fraction in Primary Cosmic Rays of 0.5--350
  GeV}},  {\em Phys. Rev. Lett.} {\bf 110} (2013) 141102.

\bibitem{Atwood:2009ez}
{\bf Fermi-LAT} Collaboration, W.~Atwood et~al., {\it {The Large Area Telescope
  on the Fermi Gamma-ray Space Telescope Mission}},  {\em Astrophys.J.} {\bf
  697} (2009) 1071--1102, [\href{http://arxiv.org/abs/0902.1089}{{\tt
  arXiv:0902.1089}}].

\bibitem{Goodenough:2009gk}
L.~Goodenough and D.~Hooper, {\it {Possible Evidence For Dark Matter
  Annihilation In The Inner Milky Way From The Fermi Gamma Ray Space
  Telescope}},  \href{http://arxiv.org/abs/0910.2998}{{\tt arXiv:0910.2998}}.

\bibitem{Hooper:2010mq}
D.~Hooper and L.~Goodenough, {\it {Dark Matter Annihilation in The Galactic
  Center As Seen by the Fermi Gamma Ray Space Telescope}},  {\em Phys.Lett.}
  {\bf B697} (2011) 412--428, [\href{http://arxiv.org/abs/1010.2752}{{\tt
  arXiv:1010.2752}}].

\bibitem{Abazajian:2010zy}
K.~N. Abazajian, {\it {The Consistency of Fermi-LAT Observations of the
  Galactic Center with a Millisecond Pulsar Population in the Central Stellar
  Cluster}},  {\em JCAP} {\bf 1103} (2011) 010,
  [\href{http://arxiv.org/abs/1011.4275}{{\tt arXiv:1011.4275}}].

\bibitem{Boyarsky:2010dr}
A.~Boyarsky, D.~Malyshev, and O.~Ruchayskiy, {\it {A comment on the emission
  from the Galactic Center as seen by the Fermi telescope}},  {\em Phys.Lett.}
  {\bf B705} (2011) 165--169, [\href{http://arxiv.org/abs/1012.5839}{{\tt
  arXiv:1012.5839}}].

\bibitem{Hooper:2011ti}
D.~Hooper and T.~Linden, {\it {On The Origin Of The Gamma Rays From The
  Galactic Center}},  {\em Phys.Rev.} {\bf D84} (2011) 123005,
  [\href{http://arxiv.org/abs/1110.0006}{{\tt arXiv:1110.0006}}].

\bibitem{Abazajian:2012pn}
K.~N. Abazajian and M.~Kaplinghat, {\it {Detection of a Gamma-Ray Source in the
  Galactic Center Consistent with Extended Emission from Dark Matter
  Annihilation and Concentrated Astrophysical Emission}},  {\em Phys.Rev.} {\bf
  D86} (2012) 083511, [\href{http://arxiv.org/abs/1207.6047}{{\tt
  arXiv:1207.6047}}].

\bibitem{Gordon:2013vta}
C.~Gordon and O.~Macias, {\it {Dark Matter and Pulsar Model Constraints from
  Galactic Center Fermi-LAT Gamma Ray Observations}},  {\em Phys.Rev.} {\bf
  D88} (2013), no.~8 083521, [\href{http://arxiv.org/abs/1306.5725}{{\tt
  arXiv:1306.5725}}].

\bibitem{Macias:2013vya}
O.~Macias and C.~Gordon, {\it {Contribution of cosmic rays interacting with
  molecular clouds to the Galactic Center gamma-ray excess}},  {\em Phys.Rev.}
  {\bf D89} (2014), no.~6 063515, [\href{http://arxiv.org/abs/1312.6671}{{\tt
  arXiv:1312.6671}}].

\bibitem{Abazajian:2014fta}
K.~N. Abazajian, N.~Canac, S.~Horiuchi, and M.~Kaplinghat, {\it {Astrophysical
  and Dark Matter Interpretations of Extended Gamma-Ray Emission from the
  Galactic Center}},  {\em Phys.Rev.} {\bf D90} (2014), no.~2 023526,
  [\href{http://arxiv.org/abs/1402.4090}{{\tt arXiv:1402.4090}}].

\bibitem{Daylan:2014rsa}
T.~Daylan, D.~P. Finkbeiner, D.~Hooper, T.~Linden, S.~K.~N. Portillo, et~al.,
  {\it {The Characterization of the Gamma-Ray Signal from the Central Milky
  Way: A Compelling Case for Annihilating Dark Matter}},
  \href{http://arxiv.org/abs/1402.6703}{{\tt arXiv:1402.6703}}.

\bibitem{Lacroix:2014eea}
T.~Lacroix, C.~Boehm, and J.~Silk, {\it {Fitting the Fermi-LAT GeV excess: On
  the importance of including the propagation of electrons from dark matter}},
  {\em Phys.Rev.} {\bf D90} (2014), no.~4 043508,
  [\href{http://arxiv.org/abs/1403.1987}{{\tt arXiv:1403.1987}}].

\bibitem{Bringmann:2012vr}
T.~Bringmann, X.~Huang, A.~Ibarra, S.~Vogl, and C.~Weniger, {\it {Fermi LAT
  Search for Internal Bremsstrahlung Signatures from Dark Matter
  Annihilation}},  {\em JCAP} {\bf 1207} (2012) 054,
  [\href{http://arxiv.org/abs/1203.1312}{{\tt arXiv:1203.1312}}].

\bibitem{Weniger:2012tx}
C.~Weniger, {\it {A Tentative Gamma-Ray Line from Dark Matter Annihilation at
  the Fermi Large Area Telescope}},  {\em JCAP} {\bf 1208} (2012) 007,
  [\href{http://arxiv.org/abs/1204.2797}{{\tt arXiv:1204.2797}}].

\bibitem{Tempel:2012ey}
E.~Tempel, A.~Hektor, and M.~Raidal, {\it {Fermi 130 GeV gamma-ray excess and
  dark matter annihilation in sub-haloes and in the Galactic centre}},  {\em
  JCAP} {\bf 1209} (2012) 032, [\href{http://arxiv.org/abs/1205.1045}{{\tt
  arXiv:1205.1045}}]. [Addendum: JCAP1211,A01(2012)].

\bibitem{Su:2012ft}
M.~Su and D.~P. Finkbeiner, {\it {Strong Evidence for Gamma-ray Line Emission
  from the Inner Galaxy}},  \href{http://arxiv.org/abs/1206.1616}{{\tt
  arXiv:1206.1616}}.

\bibitem{Yuan:2014rca}
Q.~Yuan and B.~Zhang, {\it {Millisecond pulsar interpretation of the Galactic
  center gamma-ray excess}},  {\em JHEAp} {\bf 3-4} (2014) 1--8,
  [\href{http://arxiv.org/abs/1404.2318}{{\tt arXiv:1404.2318}}].

\bibitem{Petrovic:2014uda}
J.~Petrovic, P.~D. Serpico, and G.~Zaharijas, {\it {Galactic Center gamma-ray
  "excess" from an active past of the Galactic Centre?}},  {\em JCAP} {\bf
  1410} (2014), no.~10 052, [\href{http://arxiv.org/abs/1405.7928}{{\tt
  arXiv:1405.7928}}].

\bibitem{Boehm:2014hva}
C.~Boehm, M.~J. Dolan, C.~McCabe, M.~Spannowsky, and C.~J. Wallace, {\it
  {Extended gamma-ray emission from Coy Dark Matter}},  {\em JCAP} {\bf 1405}
  (2014) 009, [\href{http://arxiv.org/abs/1401.6458}{{\tt arXiv:1401.6458}}].

\bibitem{Arina:2014yna}
C.~Arina, E.~Del~Nobile, and P.~Panci, {\it {Dark Matter with
  Pseudoscalar-Mediated Interactions Explains the DAMA Signal and the Galactic
  Center Excess}},  {\em Phys. Rev. Lett.} {\bf 114} (2015) 011301,
  [\href{http://arxiv.org/abs/1406.5542}{{\tt arXiv:1406.5542}}].

\bibitem{Hektor:2014kga}
A.~Hektor and L.~Marzola, {\it {Coy Dark Matter and the anomalous magnetic
  moment}},  {\em Phys. Rev.} {\bf D90} (2014), no.~5 053007,
  [\href{http://arxiv.org/abs/1403.3401}{{\tt arXiv:1403.3401}}].

\bibitem{Ackermann:2015zua}
{\bf Fermi-LAT} Collaboration, M.~Ackermann et~al., {\it {Searching for Dark
  Matter Annihilation from Milky Way Dwarf Spheroidal Galaxies with Six Years
  of Fermi Large Area Telescope Data}},  {\em Phys. Rev. Lett.} {\bf 115}
  (2015), no.~23 231301, [\href{http://arxiv.org/abs/1503.02641}{{\tt
  arXiv:1503.02641}}].

\bibitem{Ackermann:2015lka}
{\bf Fermi-LAT} Collaboration, M.~Ackermann et~al., {\it {Updated search for
  spectral lines from Galactic dark matter interactions with pass 8 data from
  the Fermi Large Area Telescope}},  {\em Phys. Rev.} {\bf D91} (2015), no.~12
  122002, [\href{http://arxiv.org/abs/1506.00013}{{\tt arXiv:1506.00013}}].

\bibitem{Liang:2016pvm}
Y.-F. Liang, Z.-Q. Shen, X.~Li, Y.-Z. Fan, X.~Huang, S.-J. Lei, L.~Feng, E.-W.
  Liang, and J.~Chang, {\it {Search for a gamma-ray line feature from a group
  of nearby galaxy clusters with Fermi LAT Pass 8 data}},  {\em Phys. Rev.}
  {\bf D93} (2016), no.~10 103525, [\href{http://arxiv.org/abs/1602.06527}{{\tt
  arXiv:1602.06527}}].

\bibitem{Anderson:2015dpc}
B.~Anderson, S.~Zimmer, J.~Conrad, M.~Gustafsson, M.~Sánchez-Conde, and
  R.~Caputo, {\it {Search for Gamma-Ray Lines towards Galaxy Clusters with the
  Fermi-LAT}},  {\em JCAP} {\bf 1602} (2016), no.~02 026,
  [\href{http://arxiv.org/abs/1511.00014}{{\tt arXiv:1511.00014}}].

\bibitem{Li:2012qg}
Y.~Li and Q.~Yuan, {\it {Testing the 130 GeV gamma-ray line with high energy
  resolution detectors}},  {\em Phys. Lett.} {\bf B715} (2012) 35--37,
  [\href{http://arxiv.org/abs/1206.2241}{{\tt arXiv:1206.2241}}].

\bibitem{Lu:2013kda}
T.-S. Lu, T.-K. Dong, and J.~Wu, {\it {The spatial distribution of dark matter
  annihilation originating from a gamma-ray line signal}},  {\em Res. Astron.
  Astrophys.} {\bf 14} (2014) 520--526,
  [\href{http://arxiv.org/abs/1312.0357}{{\tt arXiv:1312.0357}}].

\bibitem{ChangJin:550}
C.~Jin, {\it Dark matter particle explorer: The first chinese cosmic ray and
  hard gamma-ray detector in space},  {\em Chin. J. Space Sci.} {\bf 34}
  (2014), no.~5 550.

\bibitem{Gargano:2017avj}
{\bf DAMPE} Collaboration, F.~Gargano, {\it {DAMPE space mission: first data}},
   in {\em {25th European Cosmic Ray Symposium (ECRS 2016) Turin, Italy,
  September 04-09, 2016}}, 2017.
\newblock \href{http://arxiv.org/abs/1701.05046}{{\tt arXiv:1701.05046}}.

\bibitem{Zhang:2014qga}
{\bf HERD} Collaboration, S.~N. Zhang et~al., {\it {The high energy
  cosmic-radiation detection (HERD) facility onboard China's Space Station}},
  {\em Proc. SPIE Int. Soc. Opt. Eng.} {\bf 9144} (2014) 91440X,
  [\href{http://arxiv.org/abs/1407.4866}{{\tt arXiv:1407.4866}}].

\bibitem{2013AdSpR..51..297G}
A.~M. {Galper}, O.~{Adriani}, R.~L. {Aptekar}, I.~V. {Arkhangelskaja}, A.~I.
  {Arkhangelskiy}, M.~{Boezio}, V.~{Bonvicini}, K.~A. {Boyarchuk}, Y.~V.
  {Gusakov}, M.~O. {Farber}, M.~I. {Fradkin}, V.~A. {Kachanov}, V.~A. {Kaplin},
  M.~D. {Kheymits}, A.~A. {Leonov}, F.~{Longo}, P.~{Maestro}, P.~{Marrocchesi},
  E.~P. {Mazets}, E.~{Mocchiutti}, A.~A. {Moiseev}, N.~{Mori}, I.~{Moskalenko},
  P.~Y. {Naumov}, P.~{Papini}, P.~{Picozza}, V.~G. {Rodin}, M.~F. {Runtso},
  R.~{Sparvoli}, P.~{Spillantini}, S.~I. {Suchkov}, M.~{Tavani}, N.~P.
  {Topchiev}, A.~{Vacchi}, E.~{Vannuccini}, Y.~T. {Yurkin}, N.~{Zampa}, and
  V.~G. {Zverev}, {\it {Status of the GAMMA-400 project}},  {\em Advances in
  Space Research} {\bf 51} (Jan., 2013) 297--300,
  [\href{http://arxiv.org/abs/1201.2490}{{\tt arXiv:1201.2490}}].

\bibitem{DAmbrosio:2002ex}
G.~D'Ambrosio, G.~Giudice, G.~Isidori, and A.~Strumia, {\it Minimal flavor
  violation: An effective field theory approach},  {\em Nucl.Phys.} {\bf B645}
  (2002) 155--187.

\bibitem{Djouadi:2005gi}
A.~Djouadi, {\it {The Anatomy of electro-weak symmetry breaking. I: The Higgs
  boson in the standard model}},  {\em Phys. Rept.} {\bf 457} (2008) 1--216,
  [\href{http://arxiv.org/abs/hep-ph/0503172}{{\tt hep-ph/0503172}}].

\bibitem{Djouadi:2005gj}
A.~Djouadi, {\it {The Anatomy of electro-weak symmetry breaking. II. The Higgs
  bosons in the minimal supersymmetric model}},  {\em Phys. Rept.} {\bf 459}
  (2008) 1--241, [\href{http://arxiv.org/abs/hep-ph/0503173}{{\tt
  hep-ph/0503173}}].

\bibitem{Dolan:2014ska}
M.~J. Dolan, F.~Kahlhoefer, C.~McCabe, and K.~Schmidt-Hoberg, {\it {A taste of
  dark matter: Flavour constraints on pseudoscalar mediators}},  {\em JHEP}
  {\bf 03} (2015) 171, [\href{http://arxiv.org/abs/1412.5174}{{\tt
  arXiv:1412.5174}}]. [Erratum: JHEP07,103(2015)].

\bibitem{Calore:2014nla}
F.~Calore, I.~Cholis, C.~McCabe, and C.~Weniger, {\it {A Tale of Tails: Dark
  Matter Interpretations of the Fermi GeV Excess in Light of Background Model
  Systematics}},  {\em Phys. Rev.} {\bf D91} (2015), no.~6 063003,
  [\href{http://arxiv.org/abs/1411.4647}{{\tt arXiv:1411.4647}}].

\bibitem{Hektor:2015zba}
A.~Hektor, K.~Kannike, and L.~Marzola, {\it {Muon g-2 and Galactic Centre
  gamma-ray excess in a scalar extension of the 2HDM type-X}},  {\em JCAP} {\bf
  1510} (2015), no.~10 025, [\href{http://arxiv.org/abs/1507.05096}{{\tt
  arXiv:1507.05096}}].

\bibitem{Belanger:2010pz}
G.~Belanger, F.~Boudjema, A.~Pukhov, and A.~Semenov, {\it {micrOMEGAs: A Tool
  for dark matter studies}},  {\em Nuovo Cim.} {\bf C033N2} (2010) 111--116,
  [\href{http://arxiv.org/abs/1005.4133}{{\tt arXiv:1005.4133}}].

\bibitem{Belanger:2013oya}
G.~Belanger, F.~Boudjema, A.~Pukhov, and A.~Semenov, {\it {micrOMEGAs 3: A
  program for calculating dark matter observables}},  {\em Comput. Phys.
  Commun.} {\bf 185} (2014) 960--985,
  [\href{http://arxiv.org/abs/1305.0237}{{\tt arXiv:1305.0237}}].

\bibitem{Ade:2015xua}
{\bf Planck} Collaboration, P.~A.~R. Ade et~al., {\it {Planck 2015 results.
  XIII. Cosmological parameters}},  \href{http://arxiv.org/abs/1502.01589}{{\tt
  arXiv:1502.01589}}.

\bibitem{Bergstrom:2012vd}
L.~Bergstrom, G.~Bertone, J.~Conrad, C.~Farnier, and C.~Weniger, {\it
  {Investigating Gamma-Ray Lines from Dark Matter with Future Observatories}},
  {\em JCAP} {\bf 1211} (2012) 025, [\href{http://arxiv.org/abs/1207.6773}{{\tt
  arXiv:1207.6773}}].

\bibitem{2013AIPC.1516..288G}
A.~M. {Galper}, O.~{Adriani}, R.~L. {Aptekar}, I.~V. {Arkhangelskaja}, A.~I.
  {Arkhangelskiy}, M.~{Boezio}, V.~{Bonvicini}, K.~A. {Boyarchuk}, M.~I.
  {Fradkin}, Y.~V. {Gusakov}, V.~A. {Kaplin}, V.~A. {Kachanov}, M.~D.
  {Kheymits}, A.~A. {Leonov}, F.~{Longo}, E.~P. {Mazets}, P.~{Maestro},
  P.~{Marrocchesi}, I.~A. {Mereminskiy}, V.~V. {Mikhailov}, A.~A. {Moiseev},
  E.~{Mocchiutti}, N.~{Mori}, I.~V. {Moskalenko}, P.~Y. {Naumov}, P.~{Papini},
  P.~{Picozza}, V.~G. {Rodin}, M.~F. {Runtso}, R.~{Sparvoli}, P.~{Spillantini},
  S.~I. {Suchkov}, M.~{Tavani}, N.~P. {Topchiev}, A.~{Vacchi}, E.~{Vannuccini},
  Y.~T. {Yurkin}, N.~{Zampa}, V.~G. {Zverev}, and V.~N. {Zirakashvili}, {\it
  {Design and performance of the GAMMA-400 gamma-ray telescope for dark matter
  searches}},  in {\em American Institute of Physics Conference Series} (J.~F.
  {Ormes}, ed.), vol.~1516 of {\em American Institute of Physics Conference
  Series}, pp.~288--292, Feb., 2013.
\newblock \href{http://arxiv.org/abs/1210.1457}{{\tt arXiv:1210.1457}}.

\bibitem{Charles:2016pgz}
{\bf Fermi-LAT} Collaboration, E.~Charles et~al., {\it {Sensitivity Projections
  for Dark Matter Searches with the Fermi Large Area Telescope}},  {\em Phys.
  Rept.} {\bf 636} (2016) 1--46, [\href{http://arxiv.org/abs/1605.02016}{{\tt
  arXiv:1605.02016}}].

\end{thebibliography}\endgroup

\end{document}